\documentclass[english,aps, tightenlines,showpacs, showkeys, notitlepage]{revtex4-1}
\usepackage[latin9]{inputenc}
\usepackage{textcomp}
\usepackage{amsmath}
\usepackage{amssymb}

\makeatletter

\usepackage{dcolumn}
\usepackage{bm}

\makeatother

\usepackage{babel}
\begin{document}

\title{Friedmann equations for deformed entropic gravity}

\author{Salih Kibaro\u{g}lu$^{1}$}
\email{salihkibaroglu@gmail.com}

\author{Mustafa Senay$^{2}$}
\email{mustafasenay86@hotmail.com}

\date{\today}
\begin{abstract}
In this study, we investigate the effects of the one and two-parameters
deformed systems on the Friedmann equations of the Friedmann-Robertson-Walker
(FRW) universe in the context of the entropic gravity approach. We
give simplified forms for the deformed Unruh temperature and Einstein
field equations for three different deformed systems. Based on these
compact equations, we derive the Friedmann equations with the effective
gravitational and cosmological constant.
\end{abstract}

\affiliation{$^{1}$Department of Physics, Kocaeli University, 41380 Kocaeli,
Turkey }

\affiliation{$^{2}$Naval Academy, National Defence University, 34940 Istanbul,
Turkey}


\maketitle

\section{Introduction}

Developments on the astronomical observation in the last century show
that our universe has a form with homogeneous and isotropic characteristic
in a large scale ($10^{8}$ light years and more). This is one of
the most important properties of the universe which is called as the
cosmological principle. By the help of this principle, we suppose
that the universe has a symmetry which the curvature of the space
to be same in everywhere. In this condition, we can use the Friedmann
equations, which is derived by using Einstein's general theory of
relativity, to explain the dynamical evolution of the universe. Furthermore,
the expansion rate of the universe is accelerating which was discovered
in 1998 \citep{perlmutter1999,riess1998}. In order to formulate this
acceleration, one can add the cosmological constant and/or put an
extra energy source term, which can be viewed as the dark energy \citep{frieman2008,padmanabhan2009},
to the Einstein field equations. This generalization leads to generalize
the standard Friedmann equations with extra driving terms. From this
point of view, various cosmological models were proposed to explain
the acceleration \citep{ryden2002,mukhanov2005,weinberg2008,ellis2012}.
Thus, one can say that the extended gravity theories may take a crucial
role to explain this feature of the universe.

Recently, it has been proposed by Verlinde \citep{verlinde2011} (see
also \citep{verlinde2017}) that gravity may have a thermodynamics
origin and it can be considered as an entropic force which is called
the entropic gravity (further developments on the thermodynamical
description of gravity can also be found in \citep{jacobson1995,Padmanabhan2010,carroll2016,jacobson2016}).
Verlinde obtained the time-time components of Einstein's field equations
by motivating from the Bekenstein-Hawking entropy-area relation for
black holes \citep{bekenstein1973,hawking1974,hawking1975} and the
holographic principles. The entropic gravity also provided a useful
background to obtain generalized versions of gravity theory. Because
if one uses the generalized forms of entropy, temperature or active
gravitational mass functions instead of the standard ones, these idea
may lead to generalize the gravitational field equations. From these
ideas, many authors have extensively studied the extended gravity
theories based on the entropic gravity proposal such as \citep{ho2010,sheykhi2012,hendi2012,wang2014,moradpour2015,dil2015,Dil2017,senay2018,Kibaroglu2018,Kibaroglu2019,scardigli2018}.

In cosmological context, the standard Friedmann equations were first
derived in \citep{cai2010} by the help of Verlinde's entropic gravity
approximation and then various cosmological applications were studied
in \citep{wei2010,sheykhi2010,sheykhi2011,sheykhi2018,Sefiedgar2017,tu2018}.
Besides, in studies \citep{easson2010,easson2012}, the authors tried
to explain the accelerating universe by adding extra entropic terms,
instead of the cosmological constant, to the Friedmann equations.
They also supposed that the horizon of the universe has an entropy
and temperature. This model was named as the entropic cosmology and
it has been studied extensively in \citep{cai2010C,cai2011,koivisto2011,Komatsu2013A,Komatsu2013}.

It has been suggested that by Strominger \citep{Strominger1993} that
quantum black holes obey deformed Bose or Fermi statistics instead
of standard Bose or Fermi statistics. According to Strominger's suggestion,
the quantum black holes can be assumed $q$-deformed bosons or fermions.
Thermodynamics and statistical properties of such $q$-deformed bosons
and fermions have been extensively studied by several authors in the
literature \citep{Lavagno2000,Cai2007,Gavrilik2012,Algin2017,Mohammadzadeh2017}.
Moreover, these $q$-deformed bosons and fermions have a variety of
applications in many branches of physics. For instance, in the Ref.
\citep{dil2015,Dil2017,senay2018,Kibaroglu2018}, one and two-parameters
deformed Einstein equations were obtained by considering the quantum
black holes as deformed bosons or fermions. 

With the above motivations, in this paper, we aim to obtain deformed
Friedmann equations by using the results of three different deformed
models and we will work in a frame where $c=k_{B}=1$ where $c$ is
the speed of light and $k_{B}$ is Boltzman's constant. The paper
is organized as follows. In Sec. 2, we review the useful results for
our calculations of three different deformed models. We also give
deformed temperature function and Einstein's field equations in a
compact form. In Sec. 3, the modified Friedmann equations and the
density parameter are obtained. The last section concludes the paper
and contains some discussion on our results and possible future works.

\section{one and two-parameters deformed entropic gravity}

According to the Hawking radiation \citep{hawking1974,hawking1975},
the mass of charged black holes decreases until it reaches a critical
value proportional to its charge. Therefore, resulting structure of
the charged black holes can be named as extremal (quantum) black holes
which appears to be a quantum mechanically stable object \citep{Strominger1993}.
In addition, such black holes behave like point particles and obey
the deformed statistics. So, these black holes can be seen as deformed
bosons or deformed fermions. Furthermore, Verlinde showed that the
gravity occurs as an entropic force associated with the information
on the holographic screen. From this idea, Verlinde succeeded to derive
the time-time component of the Einstein field equation by using the
equipartition rule and Tolman-Komar mass. This consideration allows
to generalize the gravity by changing the thermodynamical quantities
such as the entropy, temperature or energy functions.

In recent years, the study \citep{dil2015} showed us that the deformed
Einstein equation can be derived by using deformed statistics rather
than standard one to describe gravitational behaviors of the quantum
black holes in the framework of the entropic gravity suggestion. From
this motivation, the different deformation of Einstein's field equations
have been obtained by considering the extremal quantum black holes
as $q$-deformed fermions \citep{senay2018}, $q$-deformed bosons
and fermions \citep{Kibaroglu2018}, and $\left(q,p\right)$-deformed
fermions \citep{Dil2017} for the high-temperature limits. In this
section, we briefly review the useful results of these three models.

In the first model \citep{senay2018}, the extremal quantum black
holes were considered as deformed fermions. The oscillators algebra
of these deformed fermions was introduced in \citep{Parthasarathy1991,Viswanathan1992,Chaichian1993}
and some of the high-temperature thermostatistical properties of these
deformed fermions were investigated in \citep{Algin2012}. The deformed
entropy function of them is defined \citep{senay2018} as

\begin{equation}
S=\frac{(2\pi m)^{3/2}V}{Th^{3}}E^{5/2}\tilde{F}(z,q),\label{eq: entropy}
\end{equation}
where 

\begin{equation}
\tilde{F}(z,q)=\frac{5}{2}f_{5/2}(z,q)-f_{3/2}(z,q)\ln z,
\end{equation}

\begin{equation}
f_{n}(z,q)=\frac{1}{|\ln q|}\left[\sum_{l=1}^{\infty}(-1)^{l-1}\frac{(zq)^{l}}{l^{n+1}}-\sum_{l=1}^{\infty}(-1)^{l-1}\frac{(z)^{l}}{l^{n+1}}\right].
\end{equation}
Also, $m$ and $T$ indicate mass and temperature of deformed particles,
respectively, and $q$ is real deformation parameter in the interval
$0<q<1$. In the statistical equilibrium, the total entropy of the
system should be constant and extremal and so the variation of the
entropy goes to zero such as

\begin{equation}
\frac{d}{dx^{a}}S(E,x^{a})=0.\label{eq: entropy derivation}
\end{equation}
Eq.(\ref{eq: entropy derivation}) can also be reexpressed

\begin{equation}
\frac{\partial S}{\partial E}\frac{\partial E}{\partial x^{a}}+\frac{\partial S}{\partial x^{a}}=0.\label{eq: entropy chain rule}
\end{equation}
where $\partial E/\partial x^{a}=-F_{a}$ and $\partial S/\partial x^{a}=\nabla_{a}S=\left(-2\pi mN_{a}\right)/\hbar$.
Considering a generalized Newton potential as a function of killing
vectors $\phi=\ln\left(-\xi^{a}\xi_{a}\right)$ in general relativity,
we can write the entropic force as $F_{a}=T\nabla_{a}S=-me^{\phi}\nabla_{a}\phi$.
Using these relations and the entropy function in Eq.(\ref{eq: entropy}),
the deformed temperature function on the holographic screen can be
derived as follows

\begin{eqnarray}
T^{\left(1\right)} & = & \frac{5V}{8\sqrt{\pi}}\frac{\left(2mE\right)^{3/2}}{h^{2}}\tilde{F}\left(z,q\right)N^{a}e^{\phi}\nabla_{a}\phi,\label{eq: temp1}
\end{eqnarray}
where $e^{\phi}$ is the redshift factor and $N^{a}$ is the unit
outward pointing vector. When taking into account the Unruh temperature,

\begin{equation}
T_{U}=\frac{\hbar}{2\pi}e^{\phi}N^{a}\bigtriangledown_{a}\phi.\label{eq: unruh temp}
\end{equation}
the deformed temperature function can be written as

\begin{eqnarray}
T^{\left(1\right)} & = & \alpha^{\left(1\right)}T_{U},\label{eq: temp1-1}
\end{eqnarray}
where the parameter $\alpha^{\left(1\right)}$ is deformation contribution
of the model to the temperature function which defined as
\begin{equation}
\alpha^{\left(1\right)}:=\frac{5V\left(2\pi mE\right)^{3/2}}{2h^{3}}\tilde{F}\left(z,q\right).
\end{equation}
Now, if we define $\Psi^{\left(1\right)}:=\tilde{F}(z,q)$ and

\begin{equation}
f\left(E,m,V\right)=\frac{5V\left(2\pi mE\right)}{2h^{3}}^{3/2},
\end{equation}
the parameter $\alpha^{\left(1\right)}$ takes the compact form as
follows

\begin{equation}
\alpha^{\left(1\right)}:=f\left(E,m,V\right)\Psi^{\left(1\right)}.
\end{equation}

In the second model \citep{Kibaroglu2018}, the quantum algebraic
structures of deformed bosons and fermions were introduced in \citep{Chaichian1993,Ng1990,Lee1990}
and some of the high and low-temperature thermostatistical properties
of these deformed bosons and fermions were examined in \citep{Lavagno2002}.
The deformed entropy function of them is defined \citep{Kibaroglu2018}
as

\begin{equation}
S=\frac{(2\pi m)^{3/2}V}{Th^{3}}E^{5/2}H^{\kappa}(z,q),\label{eq: entropy 2}
\end{equation}
where 

\begin{equation}
H^{\kappa}(z,q)=\frac{5}{2}h_{5/2}^{\kappa}(z,q)-h_{3/2}^{\kappa}(z,q)\ln z,
\end{equation}

\begin{equation}
h_{n}^{\kappa}(z,q)=\frac{1}{q-q^{-1}}\left[\sum_{l=1}^{\infty}\frac{(\kappa zq^{\kappa})^{l}}{l^{n+1}}-\sum_{l=1}^{\infty}\frac{(\kappa zq^{-\kappa})^{l}}{l^{n+1}}\right].
\end{equation}
and $q$ is real deformation parameter in the interval $0<q<1$. Similar
to the derivation of Eq.(\ref{eq: temp1}), the deformed temperature
function can be obtained as,

\begin{eqnarray}
T^{\left(2\right)} & = & \alpha^{\left(2\right)}T_{U},
\end{eqnarray}
where

\begin{equation}
\alpha^{\left(2\right)}:=f\left(E,m,V\right)H^{\kappa}(z,q)=f\left(E,m,V\right)\Psi^{\left(2\right)}.
\end{equation}

In the third model \citep{Dil2017}, both of the quantum algebraic
and some of the high-temperature thermostatistical properties of deformed
fermions were studied in \citep{Algin2014}. The deformed entropy
function of them is given \citep{Dil2017} as

\begin{equation}
S=\frac{(2\pi m)^{3/2}V}{Th^{3}}E^{5/2}F(z,q,p),
\end{equation}
where,

\begin{equation}
F(z,q,p)=\frac{5}{2}f_{5/2}(z,q,p)-f_{3/2}(z,q,p)\ln z,
\end{equation}

\begin{equation}
f_{n}(z,q,p)=\frac{1}{|\ln(q^{2}/p^{2})|}\left[\sum_{l=1}^{\infty}(-1)^{l-1}\frac{(q^{2}z)^{l}}{l^{n+1}}-\sum_{l=1}^{\infty}(-1)^{l-1}\frac{(p^{2}z)^{l}}{l^{n+1}}\right].
\end{equation}
and $(q,p)$ is real deformation parameter in the interval $0<(q,p)<1$.
Therefore, the temperature function for this model can be written
as the following form,

\begin{eqnarray}
T^{\left(3\right)} & = & \alpha^{\left(3\right)}T_{U},
\end{eqnarray}
where,

\begin{equation}
\alpha^{\left(3\right)}:=f\left(E,m,V\right)F(z,q,p)=f\left(E,m,V\right)\Psi^{\left(3\right)}.
\end{equation}
Now we can define the general form of the temperature function for
these three models as,
\begin{equation}
T=T^{\left(i\right)}=\alpha^{\left(i\right)}T_{U},\label{eq: q def temp}
\end{equation}
where the upper indice $i$ represents the corresponding model and
takes the values as $i=1,2,3$.

On the other hand, the studies \citep{dil2015,Dil2017,senay2018,Kibaroglu2018}
show that the deformed Einstein field equations can be derived as
follows by using Verlinde's entropic gravity proposal \citep{verlinde2011},

\begin{equation}
R_{\mu\nu}-\frac{1}{2}g_{\mu\nu}R=8\pi G_{eff}T_{\mu\nu}.\label{eq: def einstein field-1}
\end{equation}
In addition to this, one can introduce the effective cosmological
constant to the Einstein field equations by changing the active gravitational
mass (for more detail see\citep{senay2018}),

\begin{equation}
R_{\mu\nu}-\frac{1}{2}g_{\mu\nu}R+\Lambda_{eff}=8\pi G_{eff}T_{\mu\nu},\label{eq: def einstein field}
\end{equation}
where $R_{\mu\nu}$, $R$, $g_{\mu\nu}$ and $T_{\mu\nu}$ are the
Ricci tensor, Ricci scalar, metric tensor and energy-stress tensors,
respectively. Besides, the effective forms of the cosmological constant
and gravitational constants are defined, respectively,

\begin{equation}
\Lambda_{eff}=\frac{\Lambda}{\alpha^{\left(i\right)}},\,\,\,\,\,\,\,\,\,\,G_{eff}=\frac{G}{\alpha^{\left(i\right)}}.
\end{equation}
In this approach, we just need the deformation parameter $\alpha^{\left(i\right)}$
of the corresponding deformation model to obtain the effects on Einstein's
field equations in the framework of the entropic gravity proposal.
Therefore, we can say that this simplification can be applied to similar
deformed gas models given in \citep{tristant2014,algin2015}.

\section{Modified friedmann equations}

The Friedmann equations describe the dynamical evolution of our universe
in isotropic and homogeneous spacetime in large scales within the
framework of Einstein's theory of general relativity \citep{weinberg2008}.
It can be said that if we use generalized versions of Einstein's theory
of gravity, it may lead to modify the Friedmann equations. From this
motivation, we consider the deformed Einstein equations, as we discussed
in the previous section, to derive new contributions to the Friedmann
equations. For this purpose, we follow the procedure of \citep{cai2010}
where the standard Friedmann equations were obtained from the entropic
force together with the Unruh temperature and equipartition law of
energy. We first begin with the Friedmann-Robertson-Walker universe
with the metric,

\begin{equation}
ds^{2}=dt^{2}-a^{2}\left(t\right)\left(dr^{2}+r^{2}d\Omega^{2}\right),\label{eq: FRW metric}
\end{equation}
where $a\left(t\right)$ is a dimensionless arbitrary function of
time, known as the scale factor, which is related to the expansion
of the universe, and $\Omega$ denotes the line element of a unit
sphere. Considering Verlinde's paper \citep{verlinde2011}, there
is a spherical holographic screen $\mathcal{S}$ with a compact spatial
region $V$ and a compact boundary $\partial V$ which have following
physical radius,
\begin{equation}
\widetilde{r}=a\left(t\right)r.\label{eq: physical radius}
\end{equation}
The number of bits on the holographic screen is defined as \citep{cai2010}

\begin{equation}
N=\frac{A}{G\hbar},\label{eq: bits}
\end{equation}
where $A=4\pi\tilde{r}^{2}$ is the area of the screen, and considering
the equipartition law of energy, the total energy on the screen is
given as

\begin{equation}
E=\frac{1}{2}NT,\label{eq: energy 1}
\end{equation}
where $T$ is the temperature on the screen. Besides, we need energy-mass
relation,

\begin{equation}
E=M,\label{eq: energy 2}
\end{equation}
where $M$ corresponds to the mass in the spatial region $V$. Due
to supposing our universe have homogeneity and isotropic form in a
large scale, the matter content of the universe can be interpreted
as a perfect fluid with following stress-energy tensor,

\begin{equation}
T_{\mu\nu}=\left(\rho+p\right)u_{\mu}u_{\nu}-pg_{\mu\nu},\label{eq: e-m tensor}
\end{equation}
\begin{equation}
T=T_{\,\,\,\mu}^{\mu}=\rho-3p,\label{eq: e-m trace}
\end{equation}
where $\rho\left(t\right)$ and $p\left(t\right)$ are energy density
and the pressure of cosmological fluids, respectively. Also, $u^{\mu}=\left(1,0,0,0\right)$
represents four velocity and satisfies $g_{\mu\nu}u^{\mu}u^{\nu}=1$.
Now, we can write the contiunty equation from the conservation of
the energy stress tensor, $\nabla_{\mu}T^{\mu\nu}=0$, as follows

\begin{equation}
\dot{\rho}+3H\left(\rho+p\right)=0,\label{eq: cont eq}
\end{equation}
where $H\left(t\right)=\dot{a}/a$ is the Hubble parameter which describes
the expansion rate of the universe. We also note that a dot over any
quantity, such as $\dot{\rho}$, denotes the time derivative of that
quantity. The total mass in the spatial volume $V$ can be written
as

\begin{equation}
M=\int_{V}dV\left(T_{\mu\nu}u^{\mu}u^{\nu}\right),\label{eq: mass total}
\end{equation}
considering Eq.(\ref{eq: energy 2}), the term $T_{\mu\nu}u^{\mu}u^{\nu}$
corresponds to the energy density and it can be easily found by using
Eq.(\ref{eq: e-m tensor}) and the definition of the four velocity.
Furthermore, the acceleration of radial observer, which caused by
the matter in the spatial region, with respect to fixed $r$ at the
place of the screen is

\begin{equation}
a_{r}=-\frac{d^{2}\widetilde{r}}{dt^{2}}=-\ddot{a}r,\label{eq: acceleration}
\end{equation}
and by using this acceleration, the Unruh temperature takes the following
form

\begin{equation}
T_{U}=\frac{\hbar a_{r}}{2\pi}.\label{eq: unruh ar}
\end{equation}
Using Eqs (\ref{eq: q def temp}), (\ref{eq: bits}), (\ref{eq: energy 1}),
(\ref{eq: energy 2}), (\ref{eq: mass total}), (\ref{eq: acceleration}),
(\ref{eq: unruh ar}), the area $A=4\pi\tilde{r}^{2}$ and the volume
$V=\frac{4}{3}\pi\tilde{r}^{3}$ we get,

\begin{equation}
\ddot{\frac{a}{a}}=-\frac{4\pi G_{eff}}{3}\rho.\label{eq: newton cosmology}
\end{equation}
This equation corresponds to the dynamical equations for the Newtonian
cosmology \citep{mukhanov2005} in the deformed case. To obtain the
Friedmann equations of the FRW universe for the deformed general relativity,
we use active gravitational mass $\mathcal{M}$ rather than total
mass $M$ in the spatial region $V$. In our context, the active gravitational
mass, Tolman-Komar mass, is defined as \citep{wald1984}

\begin{equation}
\mathcal{M}=2\int_{V}dV\left(T_{\mu\nu}-\frac{1}{2}Tg_{\mu\nu}+\frac{\Lambda}{8\pi G}g_{\mu\nu}\right)u^{\mu}u^{\nu},\label{eq: mass total-1}
\end{equation}
where $\Lambda$ is the cosmological constant. After some calculations,
the active gravitational mass takes the following form

\begin{equation}
\mathcal{M}=\left(\rho+3p+\frac{\Lambda}{4\pi G}\right)V.
\end{equation}
Then, using similar calculations to the derivation of Eq.(\ref{eq: newton cosmology}),
we obtain the following expression,

\begin{equation}
\ddot{\frac{a}{a}}=-\frac{4\pi G_{eff}}{3}\left(\rho+3p\right)-\frac{\Lambda_{eff}}{3}.\label{eq: acceleration eq}
\end{equation}
Thus, we obtained the deformed version of the first Friedmann equation
(or the acceleration equation) with the cosmological constant for
the dynamical evolution of the FRW universe. After multiplying $\dot{a}a$
on both sides of the last equation and using Eq.(\ref{eq: cont eq}),
and integrating, we obtain,

\begin{equation}
H^{2}+\frac{k}{a^{2}}=\frac{8\pi G_{eff}}{3}\rho-\frac{\Lambda_{eff}}{3}.\label{eq: friedman 2}
\end{equation}
This is the deformed form of the second Friedmann equation which controls
the time evolution of the FRW universe. Here, $k$ corresponds to
the integration constant and it can be interpreted as the spatial
curvature of the region $V$ in Einstein's theory of gravity. In addition,
$k=1$, $0$ and $-1$ correspond to a closed, flat, and open FRW
universe, respectively (for more detail see \citep{ellis2012}). The
last equation can be written as

\begin{equation}
H^{2}+\frac{k}{a^{2}}=\frac{8\pi G_{eff}}{3}\left(\rho-\rho_{\Lambda}\right),\label{eq: friedman 2-2}
\end{equation}
where $\rho_{\Lambda}$ can be seen an additional energy density to
the universe which is associated with the cosmological constant,

\begin{equation}
\rho_{\Lambda}:=\frac{\Lambda}{8\pi G}.\label{eq: rho lambda}
\end{equation}
Therefore, we can easily say that the deformation does not contribute
to the energy density of the universe. Considering the continuity
equation Eq.(\ref{eq: cont eq}), one can write the pressure, $p_{\Lambda}$,
which is associated with the cosmological constant, as follows \citep{ryden2002}

\begin{equation}
p_{\varLambda}=-\rho_{\Lambda}=-\frac{\Lambda}{8\pi G}.\label{eq: p lambda}
\end{equation}
In addition, using the relation $\dot{H}=\ddot{a}/a-H^{2}$ together
with Eq.(\ref{eq: acceleration eq}) and Eq.(\ref{eq: friedman 2}),
we obtain the following expression \citep{easson2012},

\begin{equation}
\dot{H}=-4\pi G_{eff}\left(\rho+p\right)+\frac{k}{a^{2}}.\label{eq: friedmann1}
\end{equation}
This equation is another Friedmann equation for the deformed case.
Also, by the help of Eq.(\ref{eq: friedman 2}), the energy density
$\rho\left(t\right)$ can be written as follows

\begin{equation}
\rho\left(t\right)=\rho_{c}\left(t\right)+\frac{3k}{8\pi G_{eff}a^{2}}+\frac{\Lambda}{8\pi G},
\end{equation}
where $\rho_{c}\left(t\right)$ is the effective critical density
and defined as

\begin{equation}
\rho_{c}\left(t\right)=\frac{3H^{2}}{8\pi G_{eff}},\label{eq: density critical}
\end{equation}
it depends on the effective gravitational constant and the given value
of the Hubble parameter. In general, we use present value of the Hubble
parameter which is named as Hubble constant $H_{0}=H\left(t_{0}\right)=\left(\dot{a}/a\right)_{t=t_{0}}$,
and the subscript $0$ represents present values. Thus, Eq.(\ref{eq: density critical})
takes following form

\begin{equation}
\rho_{c}\left(t\right)\rightarrow\rho_{c}\left(t_{0}\right)=\frac{3H_{0}^{2}}{8\pi G_{eff}}.
\end{equation}
By the help of $\rho\left(t\right)$ and $\rho_{c}\left(t\right)$,
we can write the following expression

\begin{equation}
\Omega\left(t\right)=\frac{\rho\left(t\right)}{\rho_{c}\left(t\right)}=1+\frac{k}{H^{2}a^{2}}+\frac{\Lambda_{eff}}{3H^{2}},\label{eq: density parameter}
\end{equation}
where $\Omega\left(t\right)$ represents the density parameter (or
the cosmological parameter \citep{mukhanov2005}). Furthermore, Eq.(\ref{eq: density parameter})
can alternatively be written as \citep{ellis2012}

\begin{equation}
\Omega\left(t\right)+\Omega_{k}\left(t\right)+\Omega_{\varLambda}\left(t\right)=1,
\end{equation}
where $\Omega_{k}$$\left(t\right)$ and $\Omega_{\varLambda}\left(t\right)$
represent density parameters with respect to the integral constant
and the cosmological constant, respectively, and these parameters
are defined as

\begin{equation}
\Omega_{k}=-\frac{k}{H^{2}a^{2}},\,\,\,\,\,\,\,\Omega_{\varLambda}=-\frac{\Lambda_{eff}}{3H^{2}}.
\end{equation}
The density parameter contains the information about the shape of
our universe. For instance, if we take $\Omega=1$, this model describes
a flat universe. Besides, the condition $\Omega<1$ corresponds to
an open universe and $\Omega>1$ corresponds to a closed universe.
Currently, the present value of the density parameter is close to
one $\Omega_{0}\approx1$. Considering the present-day value of the
density parameter, a relation occurs as $\Omega_{k}\approx-\Omega_{\varLambda}$
and thus the effective cosmological constant can be found as
\begin{equation}
\Lambda_{eff}\approx-\frac{3k}{a^{2}}.
\end{equation}

\section{Conclusion}

The Friedmann equations govern the evolution of our universe for a
homogeneous and isotropic space geometry. We know that our universe
obeys this geometry condition on a large scale. So, if we modify the
Friedmann equation, this may lead to important results in the cosmology.
In this work, we studied the possible effects of the one and two parameters
deformation on the Friedmann equations. For this purpose, we considered
three different deformed gas models in the high-temperature limits.
First two models \citep{senay2018,Kibaroglu2018} are related one
parameter $q$-deformed fermion and fermion-boson gas system, respectively,
and the third one \citep{Dil2017} corresponds to two-parameters $\left(q,p\right)$-deformed
of fermion gas system. We wrote the temperature function in a compact
form depending on these models in Eq.(\ref{eq: q def temp}). We also
gave the deformed form of the Einstein field equations for given deformed
gas systems with the effective gravitational and cosmological constant
in Eq.(\ref{eq: def einstein field}) which is derived from the Verlinde's
entropic gravity proposal. According to Strominger's idea, the charged
extremal quantum black holes obey deformed statistics. Thus, one can
say that the gravitational behaviors of Strominger's black holes can
be described by the deformed Einstein field equations in Eq.(\ref{eq: def einstein field}).

Moreover, we found the deformed Friedmann equations of the FRW universe
in Eqs.(\ref{eq: acceleration eq})-(\ref{eq: friedmann1}). In addition,
we also derived the density parameter in Eq.(\ref{eq: density parameter})
with the effective cosmological term which depends on corresponding
deformation parameter $\alpha^{(i)}$. The value of the density parameter
gives important information about the shape of our universe. So, we
can say that the deformed system may take a role to explain the shape
of the universe.

According to these results, we can say that there are exact contributions
come from the deformed gas models to the Friedmann equations. In a
special limit, for instance, $q\rightarrow1$ or $\left(q,p\right)\rightarrow1$,
the models return into their ideal form in which there are no interactions
between related particles \citep{huang1987}. But, since the function
$\alpha^{(i)}$ in Eq.(\ref{eq: q def temp}) does not equal to one,
our deformed equations cannot reduce to their standard forms.

Moreover, taking into account of Eqs.(\ref{eq: acceleration eq})
and (\ref{eq: friedman 2}), our results may be interpreted as the
generalization of $\Lambda CDM$ model \citep{ryden2002,weinberg2008,ellis2012}
which is used to explain the accelerated expansion of the late universe
by including extra driving terms. In standard $\Lambda CDM$ model,
the Friedmann equations are given as follows

\begin{equation}
\ddot{\frac{a}{a}}=-\frac{4\pi G}{3}\left(\rho+3p\right)+\frac{\Lambda}{3},\label{eq: acceleration eq-1}
\end{equation}

\begin{equation}
H^{2}+\frac{k}{a^{2}}=\frac{8\pi G}{3}\rho+\frac{\Lambda}{3},\label{eq: friedman 2-1}
\end{equation}
where $\Lambda$ is the cosmological constant and the driving term
$\frac{\Lambda}{3}$ is responsible for the acceleration. Therefore,
the deformed $\Lambda CDM$ model may help to improve the explanation
of the expansion of our universe.
\begin{acknowledgments}
The authors would like to thank Oktay Cebecio\u{g}lu for valuable
discussions.
\end{acknowledgments}

\end{document}